\documentstyle[preprint,aps]{revtex}
\begin{document}
\title{Minimal SU(5) Resuscitated by Long-Lived Quarks and Leptons} 
\author{P. Q. Hung}
\address{Dept. of Physics, University of Virginia, Charlottesville,
Virginia 22901}
\date{\today}
\maketitle
\begin{abstract}
The issue of gauge unification in the (non-supersymmetric) Standard
Model is reinvestigated. It is found that with just an additional
fourth generation of non-sequential and long-lived quarks and leptons, 
$SU(3) \otimes SU(2) \otimes
U(1)$ gauge couplings converge to a common point $\sim 3.5 \times
10^{15}$ GeV ($\tau_p \sim 10^{34 \pm 1}$ years). This result is due
to the non-negligible- but still perturbative-contributions of the 
top and fourth generation
Yukawa couplings to the gauge two-loop $\beta$ functions, in
contrast with the three generation case where such a contribution is
too small to play an important role in unification.

\end{abstract}

\narrowtext


It is a standard lore that present measurements of the gauge couplings
at the Z mass appear to indicate that
all three couplings actually do not converge at the
same point if they were to evolve according to the minimal Standard Model (SM)
with three generations\cite{susy}. This is somewhat problematic for the simple
idea of Grand Unification \cite{pati,GUT}, and in particular the
minimal SU(5) model\cite{GUT}. It is
also a standard lore that with low-energy supersymmetry (generically
referred to as MSSM from here on) broken 
at around 1 TeV or less,
such a unification is possible and occurs at an energy scale $
\sim 10^{16}$ GeV corresponding to a proton lifetime of
$\sim 10^{36}$ years (roughly four orders of magnitude above the current limit).
As such, the idea of Grand Unification with a desert (beyond 1 TeV) fits
snugly with low-energy supersymmetry. There is, however, a catch. The lightest
scalar in MSSM cannot be heavier than $\sim$ 150 GeV \cite{pokorski}. What
would happen if no scalar is found below, say $2 m_Z$, thus ruling out
low-energy supersymmetry (or at least the simplest version of it)? Should
one then simply give up the idea of simple unification with a desert and
entertain more complex versions with many intermediate scales, or just
give up the whole idea of Grand Unification itself? It is perhaps
worthwhile to reexamine this whole issue within the context of the
(non-supersymmetric) SM itself.

It is not entirely clear that one has exhausted all possibilities
concerning the SM. One may ask, for example, what role the mass of the
Higgs boson, $m_H$, has, especially when it is larger than $\sim$ 174 GeV.
(For a lighter Higgs boson in the presence of a heavy fermion, 
several constraints, especially from
vacuum stability, have been discussed \cite{stability}.)
It is known from previous studies that when $m_H \geq$ 174 GeV, the Higgs
quartic coupling develops Landau poles below the Planck scale $\sim 10^{19}$
GeV\cite{hung2}. For example, the Higgs boson with a mass $\sim$ 208 GeV would develop
a Landau pole at $\sim 10^{10}$ GeV. It is therefore not at all clear how
these Landau poles might influence the evolution of the gauge couplings.

One may also ask whether or not the addition of a fourth generation might
change the evolution of the gauge couplings in such a way as to unify them
again. It is well-known that the addition of an extra family does not
change the result at one-loop. However, the two-loop $\beta$ functions
for the gauge couplings contain contributions from Yukawa couplings.
With just three generations, the dominant Yukawa contribution comes
from the top quark. However, it can be seen that the top Yukawa coupling
actually {\em decreases} with energy. As a result, it practically does not
help the convergence of the three gauge couplings. With four generations
and with the fourth generation being sufficiently heavy (an issue
explored below), it turns out that all Yukawa couplings {\em grow} with 
energy. It is found that this growth can significantly affect the
evolution of the gauge couplings.
In fact, when the fourth-generation quarks and leptons
are sufficiently heavy, all Yukawa couplings (including that of the top quark)
develop Landau poles {\em below} the Planck scale. In order to make sensible
statements based strictly on the validity of preturbation theory, we shall
restrict ourselves to the range of mass where these Landau poles lie
above a few times $10^{15}$ GeV. There are two reasons for doing so. The first
one is the fact that, in order to satisfy the current lower bound on the
proton lifetime, the unification scale (if there is one) has to be larger
than $10^{15}$ GeV. The second reason is the fact that, if the three
gauge couplings were to converge at the same point- with that point being
$\sim 10^{15}$ GeV- due to the effects of the Yukawa couplings which
show up at two loops, we would like it to happen when these Yukawa
couplings are {\em still in the perturbative domain}. That does not mean, 
however, that, if the fourth generation is sufficiently massive so that
the Landau poles are {\em below} $10^{15}$ GeV, one could not have
unification. It simply means that non-perturbative methods (such
as a Higgs-Yukawa model on the lattice for example) should be used
to examine this case. Unfortunately, such a study is not available
at the present time (e.g. the difficulty with chiral fermions on
the lattice). It is for these reasons that we shall restrict ourselves 
to the mass range where the associated Landau poles would lie {\em above}
$10^{15}$ GeV.

Of particular importance to the whole idea of Grand Unification 
\cite{pati,GUT} is the feasibility
of the search for proton decay, our {\em only} direct
evidence of such a theory. The current
prediction of supersymmetric GUT for the proton lifetime is 
approximately $10^{37}$ years, which
puts it way beyond any foreseeable future search. (It
may be a moot point if there is no light Higgs below 150 GeV.) 
For the minimal SU(5)\cite{GUT}, the well-known prediction
for the proton lifetime is roughly two orders of
magnitudes lower than the current experimental lower bound of $5.5 \times 10^{32}$
years \cite{pdecay}. Is it possible that, if the proton does decay, its lifetime might be
within reach of, say, SuperKamiokande which presumably could extend its search up
to $10^{34}$ years? We would like to point out in this note that this might be possible.

In what follows, we shall distinguish two cases: the minimal SM
with three generations and one Higgs doublet (case I), and the ``non-minimal''
case with four generations and one Higgs doublet (case II). 
We shall use two-loop RG equations throughout this paper. For
case I, they are well-known and the explicit expressions can be found in
the literature \cite{twoloop1}. For the second case with four generations, we shall
write down explicitely the two-loop RG equations below \cite{twoloop2}. 
To set the notations
straight, our definition of the quartic coupling in terms of the Higgs mass
is $m_H^2 = \lambda v^2 /3$ (corresponding to $\lambda (\phi^{\dagger} \phi)^2/6$
in the Lagrangian) while the more common definition is $m_H^2 = 2\lambda v^2$.
Therefore our $\lambda $ is {\em six} times the usual $\lambda$. The RG equations given
below reflect our convention on the quartic coupling. 

We begin with the minimal SM with three generations. As mentioned earlier,
we are particularly interested in the Higgs mass range, $m_H \geq$ 174 GeV.
In particular, if we restrict ourselves to the values of $m_H$ where the
associated Landau poles would lie above $10^{15}$ GeV, we are then looking
at the range $174 GeV \leq m_H \leq 180 GeV$. Fig. 1 shows the evolution of
the three gauge couplings $g_1, g_2, g_3$ for $m_H = 174$ GeV. (The results
are practically the same for $m_H = 180$ GeV.) One clearly sees that they
{\em do not} converge to the same point, a result similar to the already
well-known one. Here, we actually include the indirect effect of the Higgs mass,
namely its effect on the top Yukawa coupling which the feeds into the 
RG equations for the gauge couplings. As one might have suspected, we have
found no effect: the three gauge couplings still do not meet. For heavier
Higgs i.e. $m_H > 180$ GeV, the Landau poles will appear below $10^{15}$ GeV.
We cannot say for sure, at least within the context of the perturbation theory,
what influence such a ``heavy'' Higgs might have on the evolution of the
gauge couplings.

We now turn to the second scenario with four generations and one Higgs
doublet. These four generations fit snugly into $\bar{5} + 10$ representations
of SU(5), except for the right-handed neutrinos which we should need if
we were to give a mass to the neutrinos. For example, this could be
incorporated into a 16-dimensional representation of SO(10) which
splits into $\bar{5} + 10 +1$ under SU(5). We could then have a pattern of
symmetry breaking like $SO(10) \rightarrow SU(5)$ for example. All four 
neutrinos can acquire a mass via the see-saw mechanism for example.
Furthermore there is no reason why the fourth one cannot be much heavier
than the other three, namely its mass could be at least half the Z 
mass. These details are however beyond the scope of this paper.

The appropriate two-loop RG equations are given by:
\begin{mathletters}
\begin{eqnarray}
16 \pi^{2} \frac{d\lambda}{dt} =&& 4 \lambda^{2} + 4 \lambda( 3 g_{t}^{2}+
6 g_{q}^{2} + 2 g_{l}^{2}-2.25 g_{2}^{2}-0.45 g_{1}^{2})\nonumber \\
&&-12( 3 g_{t}^{4} + 6 g_{q}^{4} + 2 g_{l}^{4})
+(16 \pi^{2})^{-1}[180 g_t^{6}\nonumber \\ 
&&+288 g_q^{6}+ 96 g_l^{6} -(3 g_t^{4} + 6 g_q^{4}
+ 2 g_l^{4} - 80 g_3^{2} (g_t^{2}\nonumber \\
&&+ 2 g_q^{2}))\lambda-\lambda^{2} (24 g_t^{2} + 48 g_q^{2} + 16 g_l^{2})-(52/6)
\lambda^{3}\nonumber \\
&&-192 g_3^{2}( g_t^{4} + 2 g_q^{4})]
\end{eqnarray}
\begin{eqnarray}
16 \pi^{2} \frac{d g_t^{2}}{dt} =&& g_t^{2} \{9 g_t^{2} +12 g_q^{2} + 4 g_l^{2}
-16 g_3^{2}-4.5 g_2^{2}-1.7 g_1^{2}\nonumber \\
&&(8 \pi^{2})^{-1}  [1.5 g_t^{4}-2.25 g_t^{2}(6 g_q^{2}+ 3 g_t^{2}
+ 2 g_l^{2})\nonumber \\
&&-12 g_q^{4}- (27/4) g_t^{4} - 3 g_l^{4}+ (1/6) \lambda^{2} +g_t^{2}\nonumber \\
&&(-2 \lambda + 36 g_3^{2})-(892/9) g_3^{4}] \} 
\end{eqnarray}
\begin{eqnarray}
16 \pi^{2} \frac{d g_q^{2}}{dt} =&& g_q^{2} \{6 g_t^{2} +12 g_q^{2} + 4 g_l^{2}
-16 g_3^{2}-4.5 g_2^{2}-1.7 g_1^{2}\nonumber \\
&&(8 \pi^{2})^{-1}  [3 g_q^{4}-g_q^{2}(6 g_q^{2}+ 3 g_t^{2}
+ 2 g_l^{2})\nonumber \\
&&-12 g_q^{4}- (27/4) g_t^{4} - 3 g_l^{4}+ (1/6) \lambda^{2} +g_q^{2}\nonumber \\
&&(-(8/3) \lambda + 40 g_3^{2})-(892/9) g_3^{4}] \} 
\end{eqnarray}
\begin{eqnarray}
16 \pi^{2} \frac{d g_l^{2}}{dt} =&& g_l^{2} \{6 g_t^{2} +12 g_q^{2} + 4 g_l^{2}
-4.5 (g_2^{2}+ g_1^{2})\nonumber \\
&&(8 \pi^{2})^{-1}  [3 g_q^{4}-g_q^{2}(6 g_q^{2}+ 3 g_t^{2}
+ 2 g_l^{2}) -12 g_q^{4}\nonumber \\
&&- (27/4) g_t^{4} - 3 g_l^{4}+ (1/6) \lambda^{2} -(8/3)
\lambda g_l^{2}] \} 
\end{eqnarray}
\begin{eqnarray}
16 \pi^{2} \frac{d g_1^{2}}{dt} =&&g_1^{4} \{ (163/15)+(16 \pi^{2})^{-1}[
(787/75) g_1^{2} + 6.6 g_2^{2}\nonumber \\
&&(352/15) g_3^{2}-3.4 g_t^{2}-4.4 g_q^{2}-3.6 g_l^{2}] \}
\end{eqnarray}
\begin{eqnarray}
16 \pi^{2} \frac{d g_2^{2}}{dt} =&&g_2^{4} \{ -(11/3)+(16 \pi^{2})^{-1}[
2.2 g_1^{2} + (133/3) g_2^{2}\nonumber \\
&&32 g_3^{2}-3 g_t^{2}-3 g_q^{2}-2 g_l^{2}] \}
\end{eqnarray}
\begin{eqnarray}
16 \pi^{2} \frac{d g_3^{2}}{dt} =&&g_3^{4} \{ -(34/3)+(16 \pi^{2})^{-1}[
(44/15) g_1^{2} + 12 g_2^{2}\nonumber \\
&&-(4/3) g_3^{2}-4 g_t^{2}-8 g_q^{2}] \}
\end{eqnarray}
\end{mathletters}
In the above equations, we have assumed for the fourth family, for simplicity, 
a Dirac neutrino mass and, in order to satisfy the constraints of electroweak 
precision measurements, that both quarks and leptons are degenerate $SU(2)_L$
doublets. The respective Yukawa couplings are denoted by $g_q$ and $g_l$. 
Also, in the evolution of $\lambda$ and the Yukawa couplings, we have neglected,
in the two loop terms, contributions involving 
$\tau$ and bottom Yukawa couplings as well as
as the electroweak gauge couplings, $g_1$ and $g_2$. For the range of Higgs
and heavy quark (including the top quark) masses considered in this paper,
these two-loop contributions are not important to the evolution of $\lambda$
and the Yukawa couplings.

In what follows, we shall assume that, whatever mechanism (a right-handed
neutrino in this case) that is responsible for giving a mass to at least the 4th
neutrino will not affect the evolution of the three SM gauge couplings.
Also there are reasons to believe that this 4th generation might be rather
special, distinct from the other three and having very little mixing with them.
The physics scenario behind the 4th neutrino mass might be quite unconventional.

What masses for the fourth generation are we allowed to use in our analysis?
For the quarks, the mass can even be lower than
the top quark mass. As of now, there is no strict limit on the mass of
the 4th generation quarks if the 4th family is {\em non-sequential}, i.e.
having very little mixing with the other three. As discussed in 
\cite{frampton}, the current accessible but unexplored decay length for
a long-lived heavy quark to be detected is between 100 $\mu m$ and 1 $m$. As
long as a member of the 4th generation quark doublet (e.g. the down-type quark)
decays in that range, its mass can even be lower than the top mass. The phenomenology
of a near degenerate long-lived doublet of quarks and its detection is
discussed in full length in Ref. \cite{frampton}. As for the 4th generation leptons,
we shall assume that the mass is greater than $m_Z$.

As we have stated above, we shall restrict ourselves to the mass range
of the fourth generation that will have Landau poles only above $10^{15}$
GeV. We shall require that, if there is convergence of the three gauge
couplings, it should occur when the Higgs quartic coupling and the Yukawa
couplings are still perturbative in the sense that one can neglect contributions
coming from three loop (and higher) terms to the $\beta$ functions.

Fig. 2 shows the evolution of $g^{2}_1$, $g^{2}_2$, and $g^{2}_3$ for one
particular set of masses, namely $m_Q =$ 151 GeV, $m_L=$ 95.3 GeV and
$m_t =$ 175 GeV, where $m_Q$, $m_L$ and $m_t$ denote the fourth quark, lepton 
and top masses respectively. 
Vacuum stability ($\lambda >0$) and the requirement
that $\lambda /4 \pi \sim 1$ above $10^{15}$ GeV, for the fermion masses
listed above, give a prediction for the mass for the Higgs boson
to be $m_H = 188$ GeV which is {\em larger} than
the fourth generation quark mass. Two remarks are in order here.

First, Fig. 2 shows the evolution of the gauge couplings {\em without}
taking into account the effects of the heavy particle threshold near
the unification point. For example, that threshold could come from
the {24} and {5} Higgs scalars of SU(5). In fact, as one can see
from Fig. 2, the three gauge couplings come {\em close} (to $4\%$ or less)
to each other but do not actually meet at the same point if one does
not include heavy threshold effects. By itself, within errors, it is
already a good indication of possible unification. We would like nevertheless
to discuss the issues of heavy threshold for completeness. One may
ask the following question: if we choose a scale, $M_G$, where the
uncorrected couplings $\alpha_i \equiv g^{2}_{i}/4 \pi$ ($i=1,2,3$)
are within say $4\%$ of each other, can one bring them together
after the inclusion of heavy threshold effects? As an example, let us
take the following point (last point in Fig. 2): $ln$(E/175 GeV) = 30.62
which corresponds to $M_G = 3.48 \times 10^{15}$ GeV. At this point,
one has: $\alpha_3(M_G) = 0.0278$, $\alpha_2(M_G) = 0.0273$ and
$\alpha_1(M_G) = 0.0285$. If one defines $\Delta\alpha /\alpha \equiv
(\alpha_{larger} - \alpha_{smaller})/\alpha_{larger}$, one can immediately
see that $\Delta \alpha/\alpha \approx 2\% - 4\%$. 

Let us, e.g., assume the minimal SU(5) with the
following heavy particles:$(X,Y) =(\bar{3}, 2, 5/6) + c.c. $ with
mass $M_V$, real scalars $(8, 1, 0) + (1, 3, 1) + (1,1,0)$ (belonging
to the 24-dimensional Higgs field) with mass $M_{24}$, and
the complex scalars $(3, 1, -1/3)$ (belonging to the 5-dimensional
Higgs field). (The quantum numbers are with respect to $SU(3) \otimes
SU(2) \otimes U(1)$.) The heavy threshold corrections are then\cite{susy}
\begin{mathletters}
\begin{equation}
\Delta_1 = \frac{35}{4 \pi} ln (\frac{M_G}{M_V})-\frac{1}{30 \pi}ln 
(\frac{M_G}{M_5}) + \Delta^{NRO}_1 ,
\end{equation}
\begin{equation}
\Delta_2 = -\frac{1}{6 \pi} +\frac{21}{4 \pi} ln (\frac{M_G}{M_V})-\frac{1}{6 \pi}ln 
(\frac{M_G}{M_24}) + \Delta^{NRO}_2 ,
\end{equation}
\begin{equation}
\Delta_3 =-\frac{1}{4 \pi} + \frac{7}{2 \pi} ln (\frac{M_G}{M_V})-\frac{1}{12 \pi}ln 
(\frac{M_G}{M_5}) -\frac{1}{4 \pi} ln(\frac{M_G}{M_24})+ \Delta^{NRO}_3 ,
\end{equation}
\end{mathletters}
where
\begin{equation}
\Delta^{NRO}_i = -\eta k_{i} (\frac{2}{25 \pi \alpha_{G}^{3}})^{1/2} \frac{M_G}
{M_{Planck}},
\end{equation}
with $k_{i} =1/2, 3/2, -1$ for $i=1,2,3$, is the correction coming from 
possible dimension 5 operators present between $M_G$ and $M_{Planck}$.
The magnitude of the coefficient $\eta$ is constrained to be less 
than or equal to 10. The corrected couplings can be written as
\begin{equation}
\frac{1}{\tilde{\alpha}_i (M_G)} = \frac{1}{\alpha_i (M_G)} + \Delta_i.
\end{equation}
There are of course many choices for the different mass
scales. As an example, we shall choose: $M_5 = M_G$, $M_24 = M_G$,
$M_V = 0.5 M_G$, and $\eta =10$. With this choice and taking as $\alpha_G$
the average of $\alpha_3(M_G) = 0.0278$, $\alpha_2(M_G) = 0.0273$ and
$\alpha_1(M_G) = 0.0285$, we obtain
\begin{mathletters}
\begin{equation}
\tilde{\alpha}_1 (M_G) = 0.02705,
\end{equation}
\begin{equation}
\tilde{\alpha}_2 (M_G) = 0.02662,
\end{equation} 
\begin{equation}
\tilde{\alpha}_3 (M_G) = 0.02735.
\end{equation}
\end{mathletters}
From the above values, one can say that the couplings are practically the same
with all three $\sim$ 0.027 or $1/\tilde{\alpha}_G \sim 37$. This little exercise
shows that heavy threshold effects can indeed bring about better unification.

The second remark we wish to make is the question of the validity of 
perturbation theory. The usual requirement encountered in the literature
is that $g_{t}^{2}/4 \pi$,  $g_{q}^{2}/4 \pi$.  $g_{l}^{2}/4 \pi$, and
$\lambda/ 4 \pi$ have to be less than unity. First, notice that our
definition of $\lambda$ is {\em six times greater} than the usual one which
means that, instead of $\lambda /4 \pi < 1$, we should require here
$\lambda / 24 \pi < 1$. Let us then, for the monent, adhere to this
common requirement. (We shall come back to this point below.) At the
point where we refer to as the unification point, namely $M_G =
3.48 \times 10^{15}$ GeV, we found the following values for the
Higgs quartic and Yukawa couplings: $g_{t}^{2}/4 \pi = 0.4$,
 $g_{q}^{2}/4 \pi = 0.16$,  $g_{l}^{2}/4 \pi = 0.48$, and
$\lambda / 24 \pi = 0.19$. This clearly shows that, according
to this common criterion, one is well inside the perturbative domain
when unification occurs. Although it is not needed here, one
might even relax this criterion and replace $4 \pi$ or $24 \pi$
by $16 \pi^2$. In any case, this {\em is} perturbative unification.
A side remark is in order here. For the three generation case, the
value of the top Yukawa coupling at a similar scale is  $g_{t}^{2}/4 \pi
\sim 0.016$, a factor of 20 smaller than in the four generation case.
This explains why the Yukawa couplings are important enough in this
case, but not in the three generation case, to modify the evolution 
of the gauge couplings.

An exhaustive study of different mass combinations for the fourth generation
is beyond the scope of this paper. We wish to point out however a puzzling
feature of the four generation case. As we increase the fourth generation mass,
the Landau poles move lower in energy. Eventually, perturbation theory
ceases to be valid and the question of gauge unification cannot be answered
in this context. If one naively keeps the two-loop approximation, there seems
to be an appearance of ultraviolet fixed points at rather large values of the
Higgs quartic and Yukawa couplings. If one evolves the gauge couplings with
the presence of these ultraviolet fixed points, there again is unification
at roughly similar energy scales as the example discussed previously. Since it
is not clear that one can trust this result using only the two-loop
approximation, we are currently investigating the ``Pad\'{e}'' approximation to 
the three loop $\beta$ functions but the ``large'' number of independent 
couplings makes this procedure non-trivial although we are optimistic that
it can be carried through. However, for the purpose of this paper, as stated
previously, we shall stay with our predictions coming from 
{\em perturbation theory}.

The proton partial mean lifetime as represented by
$\tau_{p \rightarrow e^{+} \pi^{0}}$ is predicted to be
$\tau_{p \rightarrow e^{+} \pi^{0}}(yr) \approx 10^{31\pm 1}(M_{G}/4.6 \times
10^{14})^{4}$. In our case, we obtain the following prediction:
$\tau_{p \rightarrow e^{+} \pi^{0}}(yr) \approx 3.28 \times 10^{34\pm 1}$.
This is comfortably larger than the current lower limit of $5.5 \times 10^{32}$
years. In addition, the prediction is not too much larger than the
current limit which means that it might be experimentally accessible in
the not-too-distant future, in contrast with the MSSM predictions. 

This scenario made a number of predictions: 1)the proton decays at an accessible
rate $\sim 10^{34 \pm 1}$ years; 2) there is a fourth generation of 
long-lived quarks and leptons with a quark mass $\sim$ 151 GeV;
3) the Higgs mass is predicted to be $\sim$ 188 GeV $> 2 m_Z$,
a value which is well suited for the ``Gold Plated''
signal $H \rightarrow l^{+}l^{-}l^{+}l^{-}$ at the LHC. All of these
features can be tested in a not-too-distant future. For example, the fourth generation
can be {\em non-sequential} and can have exceptionally long lifetimes. This
could provide a distinct signature\cite{frampton}. 

I would like to thank Paul Frampton and Gino Isidori for helpful discussions and
comments on the manuscript.
I would also like to thank the theory groups at the University of Rome ``La Sapienza''
and at the Ecole Polytechnique, Palaiseau, for the warm hospitality where part of
this work was carried out. This work is supported in parts by the US Department
of Energy under grant No. DE-A505-89ER40518.

%
%
\begin{figure}
\caption{The evolution of the SM gauge couplings squared versus ln(E/175 GeV)
for the three generation case. $g_3$,
$g_2$, and $g_1$ are the couplings of SU(3), SU(2) and U(1) respectively. 
We have used $g_3^{2}(m_t) = 1.392$, $g_2^{2}(m_t) =0.421$, and
$g_3^{2}(m_t) = 0.2114$. The Higgs mass is $m_H$ = 174 GeV.} 
\end{figure}
\begin{figure}
\caption{The evolution of the SM gauge couplings squared versus ln(E/175 GeV)
for the four generation case. $g_3$,
$g_2$, and $g_1$ are the couplings of SU(3), SU(2) and U(1) respectively. 
We have used $g_3^{2}(m_t) = 1.392$, $g_2^{2}(m_t) =0.421$, and
$g_3^{2}(m_t) = 0.2114$. Also we use $m_Q =$ 151 GeV, $m_L=$ 95.3 GeV and
$m_t =$ 175 GeV, where $m_Q$, $m_L$ and $m_t$ denote the fourth quark, lepton 
and top masses respectively.
The heavy threshold effects are not taken into
account here. They are discussed in the text and are shown to improve the
unification point.}
\end{figure}

%
%

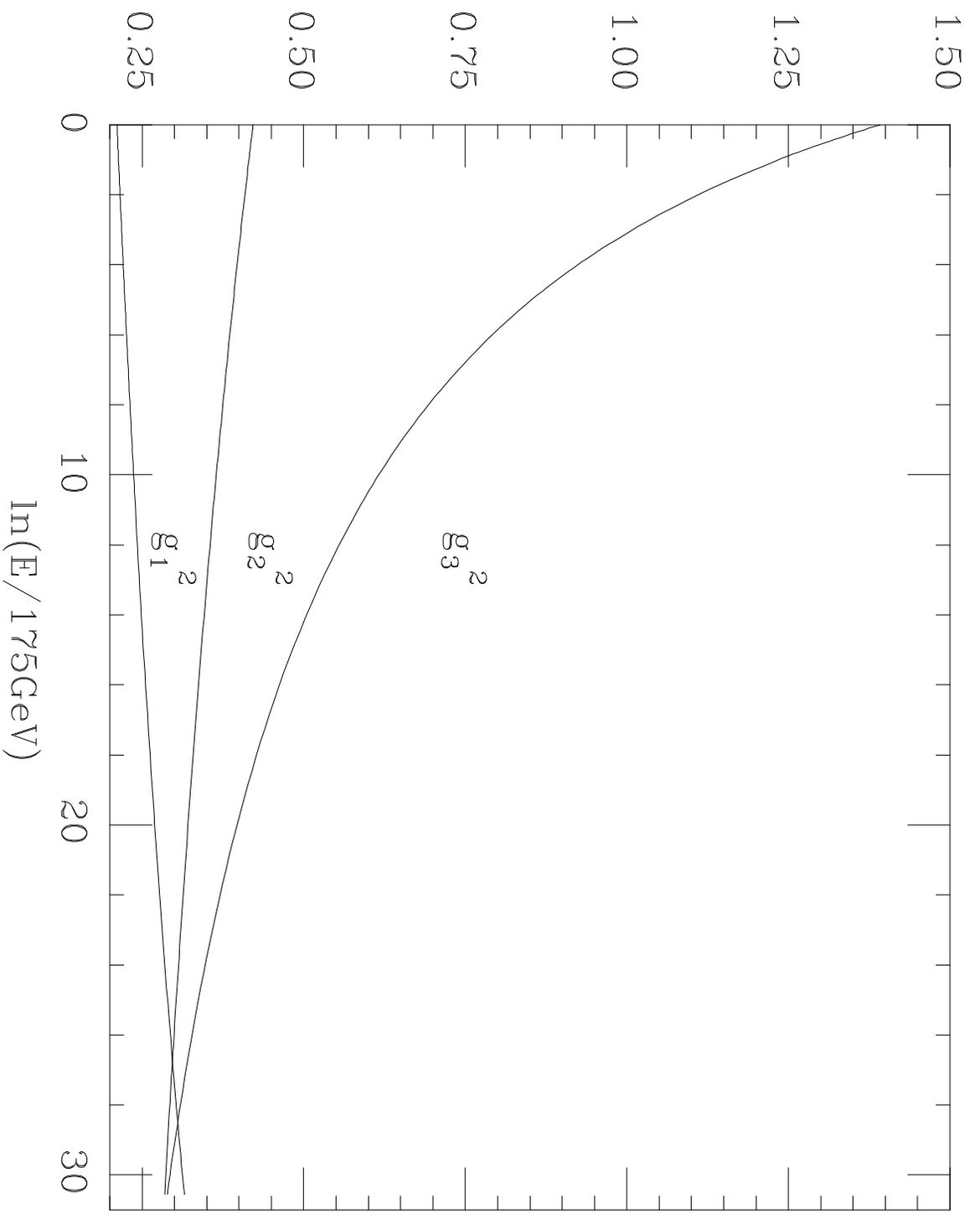
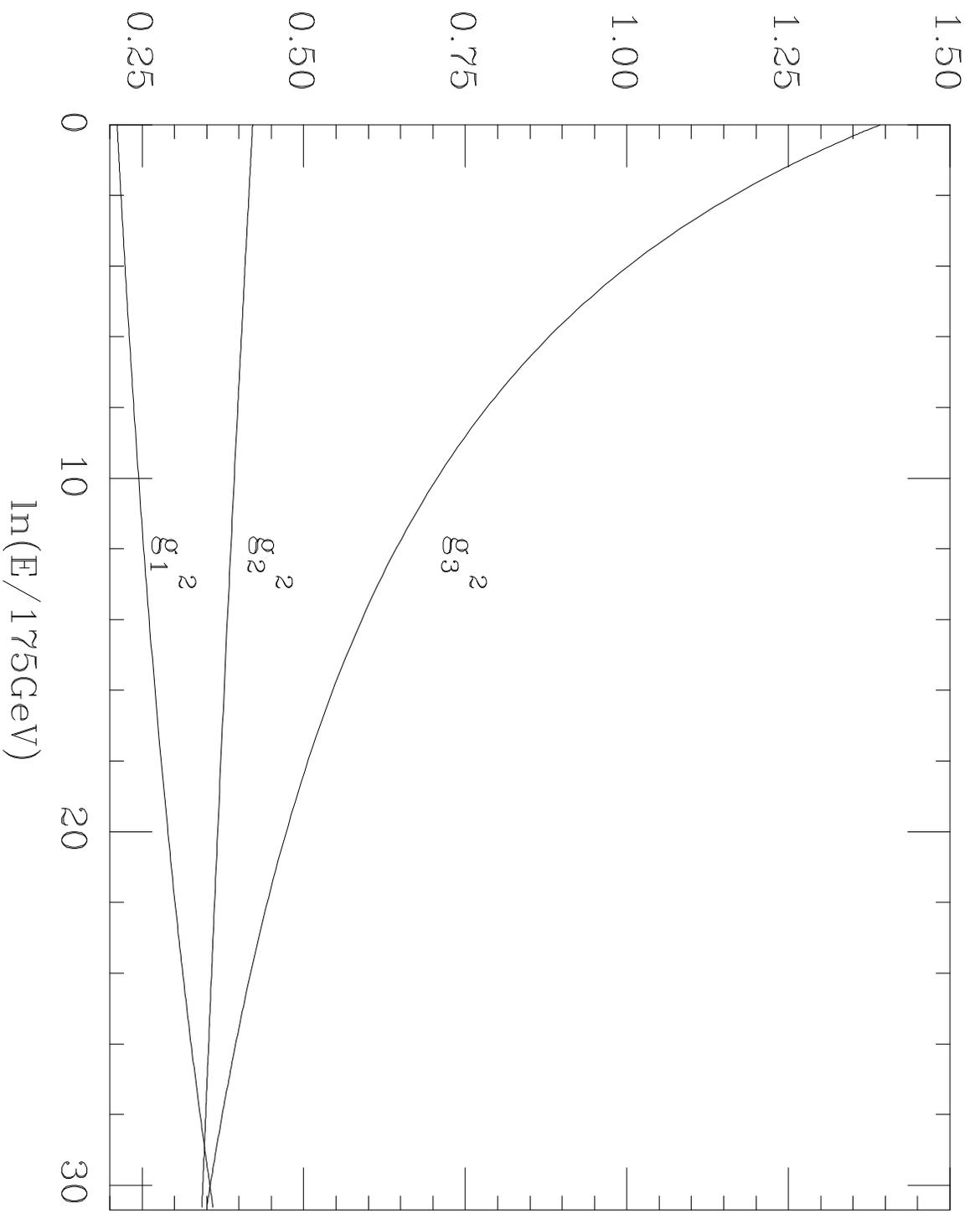
\end{document}